# Typologies of Computation viewed through the Prism of Computational Models

Mark Burgin [1], Gordana Dodig-Crnkovic [2]

[1] Department of Mathematics, UCLA, Los Angeles, USA
`mburgin@math.ucla.edu`

[2] Mälardalen University, Västerås, Sweden
`gordana.dodig-crnkovic@mdh.se`

**Abstract.** We need much better understanding of information processing and computation as its primary form. Future progress of new computational devices capable of dealing with problems of big data, internet of things, semantic web, cognitive robotics and neuroinformatics depends on the adequate models of computation. In this article we first present the current state of the art through systematisation of existing models and mechanisms, and outline basic structural framework of computation. We argue that defining computation as information processing, and given that there is no information without (physical) representation, the dynamics of information on the fundamental level is physical/ intrinsic/ natural computation. As a special case, intrinsic computation is used for designed computation in computing machinery. Intrinsic natural computation occurs on variety of levels of physical processes, containing the levels of computation of living organisms (including highly intelligent animals) as well as designed computational devices. The present article offers a typology of current models of computation and indicates future paths for the advancement of the field; both by the development of new computational models and by learning from nature how to better compute using different mechanisms of intrinsic computation.

## 1 Introduction

Many researchers have asked the question "What is computation?" trying to find a universal definition of computation or at least a plausible description of this important type of processes (cf., for example, (Turing, 1936)(Kolmogorov, 1953)(Copeland, 1996) (Burgin, 2005) (Denning, 2010)(Burgin & Dodig-Crnkovic, 2011) (Zenil, 2012)). Some did this in an informal setting based on computational and research practice, as well as on philosophical and methodological considerations. Others strived to build exact mathematical models to comprehensively describe computation. When the Turing machine (or Logical Computing Machine as Turing originally named his logical device) was constructed and accepted as a universal computational model, it was considered as the complete and exact definition of computation (Church-Turing thesis). However, the absolute nature of the Turing machine was disproved by adopting a more general definition of algorithm (Burgin, 2005).[1]

Nevertheless, in spite of all efforts, the conception of computation remains too vague and ambiguous.

This vagueness of the foundations of computing has resulted in a variety of approaches, including approaches that contradict each other. For instance, (Copeland, 1996) writes "to compute is to execute an algorithm." Active proponents of the Church-Turing Thesis, such as (Fortnow, 2012) claim computation is bounded by what Turing machines are doing (that is compute mathematical functions). For them the problem of defining computation was solved long ago with the Turing machine model. At the same time, Wegner and Goldin insist that computation is an essentially broader concept than algorithm (Goldin, Smolka, & Wegner, 2006) and propose interactive view of computing. While (Conery, 2012) argues that computation is symbol manipulation, thus disregarding analog computers computing over continuous signals, neuroscientists on the contrary study sub-symbolic computation in neurons. (Angelaki et al. 2004) Abramsky summarizes the process of successive changing of computing models as follows:

> "Traditionally, the dynamics of computing systems, their unfolding behavior in space and time has been a mere means to the end of computing the *function* which specifies the algorithmic problem which the system is solving. In much of contemporary computing, the situation is reversed: the purpose of the computing system is to exhibit certain *behaviour*. (…) We need a theory of the dynamics of informatic processes, of interaction, and information flow, as a basis for answering such fundamental questions as: What is computed? What is a process? What are the analogues to Turing completeness and universality when we are concerned with *processes* and their *behaviours, rather than the functions which they compute?* (Abramsky, 2008)

Abramsky emphasizes that there is the need for *second-generation models of computation*, and in particular process models. *The first generation models of computation* originated from problems of formalization of mathematics and logic, while processes or agents, interaction, and information flow are results of recent developments of computers and computing. In the second-generation models of computation, previous isolated systems are replaced by processes or agents for which the *interactions* with each other and with the environment are fundamental. Hewitt too advocates agent-type, Actor model of computation (Hewitt, 2012) which is suitable for modeling of physical (intrinsic) computation.

Existence of various types and kinds of computation, as well as a variety of approaches to the concept of computation, shows remarkable complexity that makes communication of results and ideas increasingly difficult. Our aim is to explicate present diversity and so call attention to the necessity of common understanding: different models of computation may have their specific uses and applications. It is just necessary to understand their mutual relationships and assumptions under which they apply.

---

[1] In the same spirit of broadening of the concept of computation, the following definition was proposed: "For a process to qualify as computation a model must exist such as algorithm, network topology, physical process or in general any mechanism which ensures definability of its behavior." (Dodig-Crnkovic, 2011)

In this paper, we present methodological analysis of the concept of computation before and after electronic computers and the emergence of computer science, demonstrating that history brings us to the conclusion that efforts in building such definitions by traditional approaches would be inefficient. An effective methodology is to find essential features of computation with the goal to explicate its nature and to build adequate models for research and technology. In addition, we perform structural analysis of the concept of computation through explication of various structures intrinsically related to computation.

To start with, we depict computation in the historical perspective, demonstrating the development of this concept on the practical level related to operations performed by people and computing devices, as well as on the theoretical level where computation is represented by abstract (mostly mathematical) models and processes. This allows us to discover basic structures inherent for computation and to develop a multifaceted typology of computations.

The paper is organized in the following way. In Section 2, we present methodological and historical analysis of the conception of computation. Section 3, we develop computational typology, which allows us to extract basic characteristics of computation and separate fundamental computational types. The suggested system of classes allows us to reflect natural structures in the set of computational processes. In Section 4, we study the structural context of computation, illuminating the Computational Triad, the Shadow Computational Triad and several other triads and pyramids intrinsically related to computation. In Section 5 we present the development of computational models, and particularly natural computing. Finally, we summarize our findings in Section 6.

## 2   Methodological and Historical Development of the Notion of Computation

If you ask nowadays who is a computer and what is computation, 99 people out of a hundred will tell that computer is not *who* but *what* because it is an electronic device, while computation is what computers are doing. However, for a long time, the term computer was associated with a human being. As Parsons and Oja (1994) write, "if you look in a dictionary printed anytime before 1940, you might be surprised to find a computer defined as a *person* who performs calculations. Although machines performed calculations too, these machines were related to as calculators, not computers."

This helps us to understand the words from the famous work of Turing (1936). Explaining first how his fundamental model, which later was called a Turing machine, works, Turing writes: "We may now construct a machine to do the work of this computer." Here a computer is a person and not a machine.[2]

Calculations of people-computers were governed by definite rules and in the medieval Europe these rules acquired the name *algorithms*. The word "*algorithm*" has an interesting

---

[2] On the topic of human computers, Burgin reports: "Even in a recently published book (Rees, 1997), we can read, "On a chi-chhou day in the fifth month of the first year of the Chih-Ho reign period (July AD 1054), Yang Wei-Te, the Chief Computer of the Calendar – the ancient Chinese counterpart, perhaps, of the English Astronomer Royal – addressed his Emperor …" It does not mean that the Emperor had an electronic device for calendar computation. A special person, who is called Computer by the author of that book, performed necessary computations." (Burgin, 2005)

historical origin. It derives from the Latin form of the name of the famous medieval mathematician Muhammad ibn Musa al-Khowarizmi (800 A.D. – 847) who wrote his main work *Al-jabr wa'l muqabala* (from which our modern word "*algebra*" is derived) and a treatise on Hindu-Arabic numerals. The Arabic text of the latter book was lost but its Latin translation, *Algoritmi de numero Indorum,* which means in English *Al-Khowarizmi on the Hindu Art of Reckoning*, introduced to the European mathematics the Hindu place-value system of numerals based on the digits 1, 2, 3, 4, 5, 6, 7, 8, 9, and 0. The first introduction to the Europeans of the use of zero as a placeholder in a positional base notation was probably also due to al-Khowarizmi in this work. Various methods for arithmetical calculation in a place-value system of numerals were given in this book as well. In the twelfth century, his works were translated from Arabic into Latin. Methods described by al-Khowarizmi were the first to be called algorithms following the title of the book. For a long time, algorithms meant the rules for people to use in making calculations. Such people were called *computers*, and their operation with numbers was also called *computation*.

Thus, algorithms were the first models of computation. Over time, the meaning of the word *algorithm* has extended as shown in (Chabert, 1999). Originating in arithmetic, algorithm was explained as the practice of algebra in the 18$^{th}$ century. In the 19$^{th}$ century, the term came to mean any process of systematic calculation. In the 20$^{th}$ century, Encyclopedia Britannica described algorithm as a systematic mathematical procedure that produces – in a finite number of steps – the answer to a question or the solution of a problem.

Historically, models of computation were first mathematical constructs as mathematicians tried to capture the notion of algorithm by rigorous formal constructions. The succeeding historical exposition is following the account from (Burgin, 2005):

The first model was λ-*calculus* built by Church (1932/33). Creating λ-calculus, Church was developing a logical theory of functions and suggested a formalization of the notion of computability by means of λ-definability. It is interesting to know that the theory of Frege (1893) actually contains λ-calculus. So, there were chances to develop a theory of algorithms and computability in the 19$^{th}$ century. However, at that time the mathematical community did not feel a need in such a theory and probably, would not accept it if somebody created it.

The next model, *recursive functions*, was introduced by Gödel (1934) in his 1934 Princeton lectures. This concept was based on ideas of Herbrand, while the construction wass equivalent to the current notion of *general recursive function*. As it is stated in (Barbin, *et al*, 1999), the structure of recursive function formalizes double recursion, while a function defined by double recursion appeared in the works of Ackermann in 1920. Hilbert presented this function in 1925 in a lecture, so as to prove the continuum hypothesis, and Ackermann studied it in 1928.

Gödel expected that all effectively computable functions are general recursive, but was not convinced of this (nor of the first version of the Church's thesis of 1934 that stated that effective computability is equivalent to λ-computability) until he became acquainted with the work of Turing (1936), in which Turing introduced and studied ordinary *Turing machines*. Similar construction was developed by Post in 1936. Besides, Church (1936) brought in *recursive functions* and Kleene (1936) presented *partial recursive functions*. Later Kleene demonstrated how λ-definability is related to the concept of recursive function, and Turing (1937) showed how λ-definability is related to the concept of Turing machine.

After this a diversity of mathematical models of algorithms has been suggested. [The following paragraph is adapted from (Burgin & Dodig-Crnkovic, 2013) in order to illustrate the vast variety of existing models of computation]: They include a variety of Turing machines: *multihead, multitape Turing machines, Turing machines with n-dimensional tapes, nondeterministic, probabilistic, alternating* and *reflexive Turing machines*, *Turing machines with oracles*, *Las Vegas Turing machines*, etc.; *neural networks* of various types – *fixed-weights, unsupervised, supervised, feedforward,* and *recurrent neural networks*; *von Neumann automata* and general *cellular automata*; *Kolmogorov algorithms finite automata* of different forms – *automata without memory, autonomous automata, automata without output or accepting automata, deterministic, nondeterministic, probabilistic automata*, etc.; *Minsky machines*; *Storage Modification Machines* or simply, *Shönhage machines*; *Random Access Machines* (RAM) and their modifications - *Random Access Machines with the Stored Program* (RASP), *Parallel Random Access Machines* (PRAM); *Petri nets* of various types – *ordinary* and ordinary *with restrictions, regular, free, colored,* and *self-modifying Petri nets*, etc.; *vector machines*; *array machines*; *multidimensional structured model of computation and computing systems*; *systolic arrays*; *hardware modification machines*; *Post productions*; *normal Markov algorithms*; *formal grammars* of many forms – *regular, context-free, context-sensitive, phrase-structure*, etc.; and so on. In addition, direct models of computational processes were introduced. [3]The variety of models of algorithms and computational processes resulted in the corresponding spectrum of types and kinds of computations, which are classified in the following section.

## 3 Computational Typologies

It is common that we talk about computation as if it would be a uniquely defined concept. However as we mentioned in the introduction, some think of computation as algorithm, others as symbol manipulation, while yet others may have in mind a more general phenomenon of information processing. Currently there are many types and kinds of computations known and used by people and it is useful to make classification and systematization of various types of computation so to better understand their mutual relationships. In what follows we will present the several main typologies of computation: existential/substantial, organizational, temporal, representational, data-oriented, operational, process-oriented and level-based.

### 3.1  Existential/substantial typology of computations

According to Burgin, (Burgin, 2012) p. 93, the basic structure of the world is represented by the existential triad (physical, structural, mental) world which corresponds to Plato's triad (material, ideas/forms, mental) world. This can also be related to Peirce's corresponding triad of (object, sign, interpretant). The existential/substantial classification of computation is based on the existential triad, and thus defines the following types of computations:

1. Physical or embodied (object) computations
2. Abstract or structural (sign) computations
3. Mental or cognitive (interpretant) computations

---

[3] Examples are the CSP model of Hoare (1985), the ACP model (Bergstra and Klop, 1984; Baeten and Bergstra, 1991), the CCS model of Milner (1989), ESP (event-signal-process) model (Lee and Sangiovanni-Vincentelli (1996), the VCR model (Smith, 2000), the EVCR (extended view-centric reasoning) model (Burgin and Smith, 2006, 2007), and the EAP (event-action-process) model (Burgin and Smith, 2010).

Abstract and mental computations are always represented by some embodied/physical/object computations.

The existential types from this typology have definite subtypes. The following are known types of physical/embodied computations.

1.1. Physical computations[4]

1.2. Chemical computations

1.3. Biological computations

In case of structural/abstract computations it is possible to discern the following types:

2.1  Symbolic computations

2.2  Subsymbolic computations

There are connections between the above types. For instance (Bucci, 1997) in the Multiple Code Theory differentiates between *verbal* (linguistic) (symbolic and subsymbolic) and non-verbal (non-linguistic) (symbolic and subsymbolic)[5] mental (cognitive) processes. Bucci suggests that the principle of object formation may be an example of the transition from a stream of massively parallel subsymbolic microfunctional events to symbol-type, serial processing through subsymbolic integration. (Clark, 1989) suggests the similar connection between symbolic and subsymbolic (connectionist) computations.

Mental (cognitive) computations can be observed at the following levels:

3.1  Individual cognitive computations

3.2  Group cognitive computations

3.3  Social cognitive computations

---

[4] In this context, currently prevailing electronic computers operate at the physical level.

[5] (Bucci, 1997) regarding natural language (of which formal languages are a subset): *The verbal symbolic domain* concerns aspects of mental functions regarding words and language. *The non-verbal symbolic domain* concerns images and representations of objects and events that do not have the form of verbal language (such as iconic visual images). *The non-verbal subsymbolic* includes experiences like intuition and the emotional communication. *The verbal subsymbolic* concerns the non-symbolic aspects of language such as prosody, meter, rhythm and the phonemic qualities of language. Sounds of words can have direct relationship to the objects they represent.

### 3.2 Organizational typology of computations

The following types of organizational topology of computation can be distinguished:

1. *Centralized computations* where computation is controlled by a single algorithm.

2. *Distributed computations* where there are separate algorithms that control computation in some neighbourhood. Usually a neighbourhood is represented by a node in the computational network. A neighbourhood can consist of a single computational device.

3. *Clustered computations* where there are separate algorithms that control computation in clusters of neighbourhoods.

Turing machines, partial recursive functions and limit Turing machines are models of centralized computations. Neural networks, Petri nets and cellular automata are models of distributed computations. Grid automata in which some nodes represent networks with the centralized control and the World Wide Web are systems that perform clustered computations (Burgin, 2005). Note that grid automaton is the most advanced abstract model of distributed computing systems, which performs concurrent computations, while being a physical distributed computing system.

### 3.3 Temporal typology of computations

With respect to temporal characteristics, computations can be characterized as:

1. *Sequential computations*, which are performed in linear time.

2. *Parallel or branching computations*, in which separate steps (operations) are synchronized in time.

3. *Concurrent computations*, which do not demand synchronization in time.

Note that while parallel computation is completely synchronized, branching computation is not completely synchronized because separate branches acquire their own time and become synchronized only in interactions.

Classical models of computation, such as the classical Turing machine or partial recursive functions, perform only sequential computations. Models that appeared later, such as Turing machines with several heads and tapes or cellular automata, provide means for parallel computations. There are also various models for concurrent computations, according to (Burgin, Mark and Smith, 2008)[6] In this context, the most advanced device model is grid automaton, while the most advanced operational model, which also is a process model, is

---

[6]"Petri nets (Petri, 1962; Reisig, 1985), Kahn process networks (Kahn, 1974), discrete event simulators (Fishman, 1978), the CSP model of Hoare (1985), the Linda model (Gelernter, 1985), the ACP model (Bergstra and Klop, 1984; Baeten and Bergstra, 1991), the Actors model (Agha, 1986), the CCS model of Milner (1989), dataflow process networks (Lee and Parks, 1995), ESP (event-signal-process) model (Lee and Sangiovanni-Vincentelli, 1996), the VCR model (Smith, 2000), grid automata (Burgin, 2005), the EVCR (extended view-centric reasoning) model (Burgin and Smith, 2006), and the EAP (event-action-process) model (Burgin and Smith, 2010)."

the EAP (event-action-process) model. All these models form three classes:

3.1  *Device models* (Petri nets, Kahn process networks, dataflow process networks, discrete event simulators, grid automata, the Linda model and the Actors model).

3.2  *Operational models* (ACP, ESP, VCR, EVCR and ESP model).

3.3  *Process models* (CSP model and CCS model).

### 3.4  Representational typology of computations

Concerning the data representation on which computations are performed, there are following types of computation:

1. *Discrete computations,* which include interval computations.

2. *Continuous computations,* which include fuzzy continuous computations.

3. *Hybrid/Mixed computations,* which include discrete and continuous processes.

Digital computing devices and the majority of computational models, such as finite automata, Turing machines, recursive functions, inductive Turing machines, and cellular automata, perform discrete computations.

Examples of continuous computations are given by abstract models, such as general dynamical systems (Bournez, 1977) and hybrid systems (Gupta, Jagadeesan, & Saraswat, 1998), and special computing devices, such as the differential analyzer (Shannon, 1941; Moore, 1996).

Hybrid/Mixed computations include piecewise continuous computations, combining both discrete computation and continuous computation. Examples of mixed computations are given by neural networks (McCulloch & Pitts, 1990), finite dimensional machines and general machines of (Blum, Cucker, Shub, & Smale, 1996).

It is possible to refine the representational typology in the following way, obtaining three additional classifications.

### 3.5  Data-oriented typology of computations

With respect to the data and the domain of computation, the following possibility exist:

1. The domain of computation is discrete and data are finite. For instance, data are words in some alphabet.

2. The domain of computation is discrete but data are infinite. For instance, data are ω-words in some alphabet. This includes interval computations because real numbers traditionally are represented as ω-words.

3. The domain of computation is continuous.

### 3.6   Operational typology of computations

With respect to operations in computation, the following typology can be identified:

1. Operations in computation are discrete and they transform discrete data elements. For instance, addition or multiplication of whole numbers.

2. Operations in computation are discrete but they transform (operate with) continuous sets. For instance, addition or multiplication of all real numbers or of real functions.

3. Operations in computation are continuous. For instance, integration of real functions.

### 3.7   Process-oriented typology of computations

1. The process of computation is discrete, i.e., it consists of separate steps in the discrete domain, and it transforms discrete data elements. For instance, computation of a Turing machine or a finite automaton.

2. The process of computation is discrete but it employs continuous operations. An example is given by analogue computations (Shannon, 1941; Scarpellini, 1963; Moore, 1996) as quoted in (Burgin, 2005) p. 122.

3. The process of computation is continuous but it employs discrete operations. For instance, computation of a limit Turing machine (Burgin, 2005) p. 140.

### 3.8   Levels of computation typology

In (Burgin & Dodig-Crnkovic, 2011) three *generality levels of computations* are introduced.

1. *The top and most abstract/general level*, where computation is perceived as any transformation of information and/or information representation.

2. *The middle level,* where computation is distinguished as a discretized process of transformation of information and/or information representation.

3. *The bottom, least general level*, where computation is recognized as a discretized process of symbolic transformation of information and/or symbolic information representation.

There are also spatial levels or scales of computations:

1. *The macrolevel* includes computations performed by mechanical calculators as well as electromechanical devices.

2. *The microlevel* includes computations performed by integrated circuits.

3. *The nanolevel* includes computations performed by fundamental parts that are not bigger than a few nanometers.

4. *The molecular level* includes computations performed by molecules.

5. *The quantum level* includes computations performed by atoms and subatomic particles.

At present there are no commercially available nanocomputers, molecular or quantum computers. However, current chips produced by nanolithography are close to computing nanodevices because their basic elements are less than 100 nanometers in scale.

## 4   Structural Framework of Computation

As presented in the historical exposition, the first idea of computation was application of an algorithm. Even Turing machine model of computation is equivalent to algorithm. Thus, the first and most commonly encountered computational structure is the Computational Dyad (cf. Figure 1), (Burgin, Mark and Eberbach, 2012)

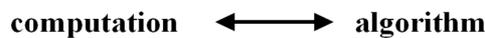

**computation** ⟷ **algorithm**

**Figure 1.** The Computational Dyad

The Computational Dyad reflects the existing duality between computations and algorithms. According to (Denning, 2010), in the 1970s Dijkstra defined an algorithm as a static description of computation, which is a dynamic state sequence evoked from a machine by the algorithm. Later a more systemic explication of the duality between computations and algorithms was elaborated. Namely, computation is a process of information transformation, which is organized and controlled by an algorithm, while an algorithm is a system of rules for a computation (Burgin, 2005). In this context, an algorithm is a compressed informational/structural representation of a process.

Note that a computer program is an algorithm written in (represented by) a programming language. Thus an algorithm is an abstract structure and it is possible to realize the same algorithm as different programs (in different programming languages). It is important to understand the difference between algorithm and its representation or embodiment. An algorithm is an abstract structure, which can be represented in a multiplicity of ways: as a computer program, a control schema, a graph, a system of cell states in the memory of a computer, a mathematical system, such as an abstract finite automaton, etc.

Interestingly, many people think that neural networks perform computations without algorithms. However, this is not true as neural networks algorithms have representations even though very different from traditional representations of algorithms as systems of rules/instructions. The neural networks algorithms are represented by neuron weights and connections between neurons. This is similar to hardware representation/realization of algorithms in computers (analog computing).

However, the Computational Dyad is incomplete because there is always a system that uses algorithms to organize and control computation. This observation shows that the Computational Dyad has to be extended to the Triangular Basic Computational Triad (Figure 2).

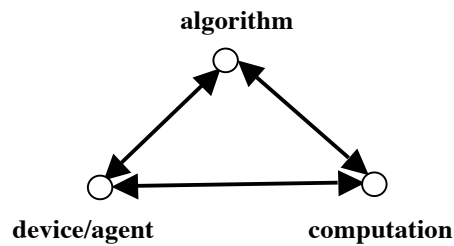

**Figure 2.** The Triangular Basic Computational Triad (TBCT)

The Triangular Basic Computational Triad (TBCT) reflects that computation is performed by a device, such as a computer, or by an agent, such as a human being or bacteria. Functioning of the device or work of the agent that/who performs computation is directed or controlled by an algorithm embodied in a program, plan, scenario or hardware. Consequently, the process of computation is also directed/controlled by an algorithm.

Note that the computing *device* can be either a *physical device*, such as a computer, or an *abstract device*, such as a Turing machine, or a *programmed* (*virtual or simulated*) *device* when a program simulates some physical or abstract device. For instance, neural networks and Turing machines are usually simulated by programs on conventional computers. A Java virtual machine can be run on different operating systems and is processor- and operating system- independent. Besides, with respect to architecture, it can be an *embracing device*, in which computation is embodied and exists as a process, or an *external device*, which organize and control computation as an external process.

The Computational Triad reflects the structure of the world represented by the Existential Triad (Burgin, 2012). The Existential Triad develops the tradition of Plato, Aristotle and Popper, and is presented in Figure 3, giving a consistent understanding and scientifically grounded interpretation of Plato ideas.

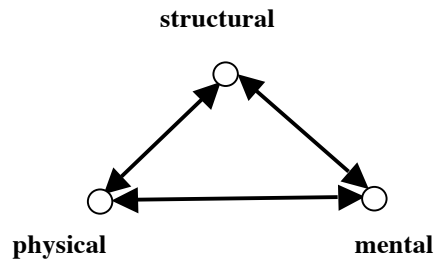

**Figure 3.** The Existential Triad of the World

Here the physical (material) world is interpreted as the physical reality studied by natural sciences, the mental world encompasses different levels of mentality (cognition), and the world of structures consists of various forms and types of structures.

The correspondence between the Computational Triad and the Existential Triad relies on the fact that computing device belongs to the physical world, algorithm is the structural representation of computation and computation itself goes in the mental (cognitive) world, which according to existential triad theory, is understood in a broader way than in the traditional interpretation of mentality (Burgin, 2012).

In addition to the components of the Computational Triad, there are other objects essentially related to computation. Computation always goes in some environment and within some context. Computation always works with data performing data transformations. Besides, it is possible to assume that computation performs some function and has some goal (for some agent) even if we don't know this goal. The basic function of computation is information processing.

These considerations bring us to a new structure called the Triangular Shadow Computational Triad (cf. Figure 4). In this triad, data come from the environment, their meaning and processing depend on the context, while the whole computation is function-oriented or goal-oriented. Computations initiated or performed by people are goal-oriented, while computational processes in nature are function-oriented.

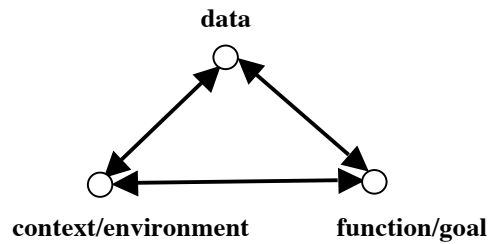

**Figure 4.** The Triangular Shadow Computational Triad (TSCT)

Thus, the Triangular Shadow Computational Triad complements the Triangular Basic Computational Triad reflecting that any computation has a goal, goes on in some context, which includes environment, and works with data. In a computation, information is processed by data transformations.

In addition, the Triangular Shadow Computational Triad is homological (in the biological sense) to the Triangular Basic Computational Triad. Namely, *data* correspond to *algorithm* as in some computational situations, data play the role of an algorithm, or more exactly, the embodiment of the algorithm in the form of a program, while the embodiment of algorithm plays the role of data. An example of this situation is (traditional) simulation testing of software when program is performed with different datasets to check the results of program functioning.

In the homological correspondence between computational triads, the triad component *context*/*environment* corresponds to the triad component *device*. More exactly, *context* corresponds to abstract devices, such as finite automata or Turing machines, while *environment* corresponds to physical devices, such as computers.

The *function/goal* component of the Triangular Shadow Computational Triad corresponds to the component *computation* of the Triangular Basic Computational Triad. More exactly, *function* corresponds to computations that go on in *nature*, while *goal* corresponds to *artificial* devices, such as computers, which are designed to achieve some goal through their computations since they are used as tools by humans.

To control computers and other computing devices, an algorithm has to be implemented into a program. This brings us to one more computational triad called the Triangular Action Computational Triad (cf. Figure 5).

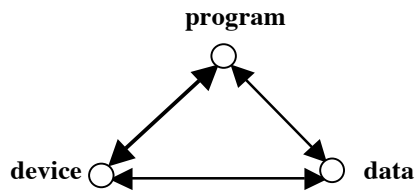

**Figure 5.** The Triangular Action Computational Triad (TACT)

In addition to triangular computational triads, there are also *synthetic* computational triads. The first such a triad (cf. Figure 6) was defined by Wirth (1975) who called his book *Algorithms + Data Structures = Programs*. In this triad, algorithms and data structures are synthesized into programs.

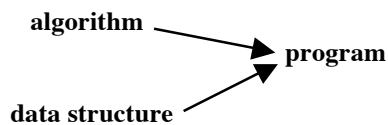

**Figure 6.** The Synthetic Computational Triad of Wirth (SCTW)

A special case of the above structure is synthetic computational triad of Michalewicz (1996) (cf. Figure 7) who titled his book *Genetic Algorithms + Data Structures = Evolution Programs*. In this triad, genetic algorithms and data structures are synthesized into evolution programs.

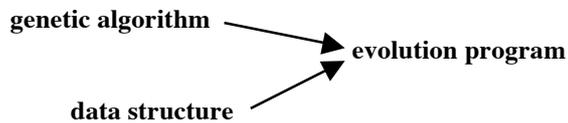

**Figure 7.** The Synthetic Computational Triad of Michalewicz (SCTM)

Here we introduce one more synthetic computational triad (cf. Figure 8), in which operations applied to the data domain generate a process in general and computation in particular.

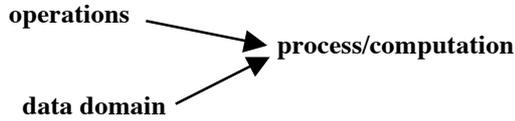

**Figure 8.** The Synthetic Computational Triad of Processes (SCTP)

All triads considered above are *flat structures*. At the same time, further development of these triads gives us *spatial structures*. Namely, computation brings a new dimension to the Triangular Action Computational Triad (TACT) transforming it into the Action Computational Pyramid (ACP) presented in Figure 9.

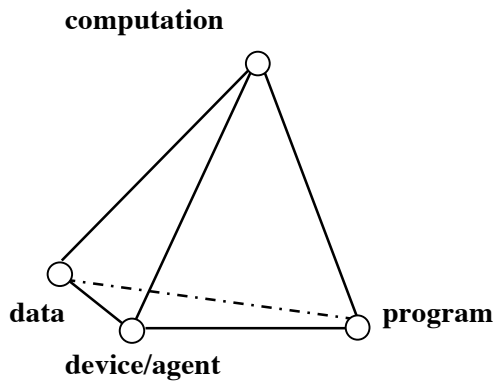

**Figure 9.** The Action Computational Pyramid (ACP)

Combination of the Triangular Action Computational Triad with the Synthetic Computational Triad of Wirth generates one more special structure called the Plain Computational Pyramid (PCP) and presented in Figure 10.

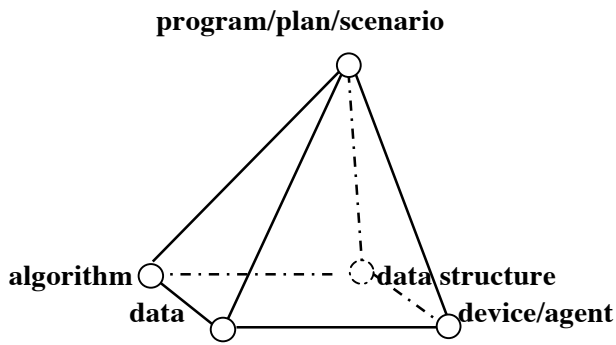

**Figure 10.** The Basic Computational Pyramid (TBCP)

Context/environment brings a new dimension to the Action Computational Pyramid (TACP) making it the Four-dimensional (Superspatial) Action Computational Pyramid (SSACP). Here we cannot draw a picture of a four-dimensional pyramid. So, we present only its three-dimensional faces in Figure 11.

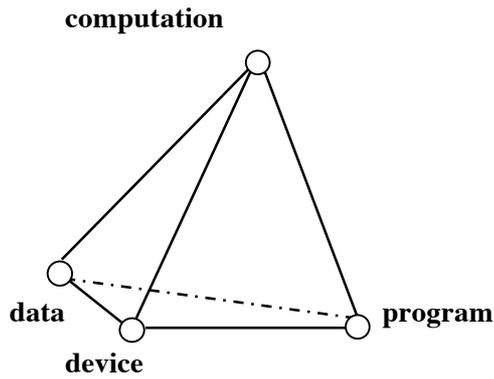

a)   The first face

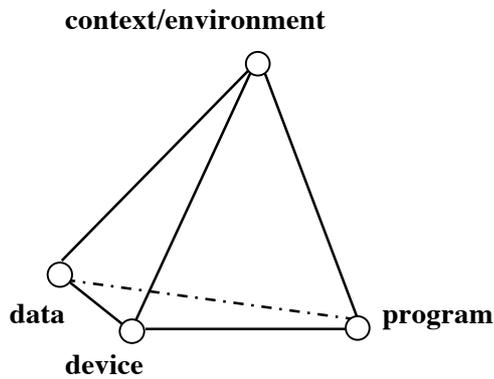

b)   The second face

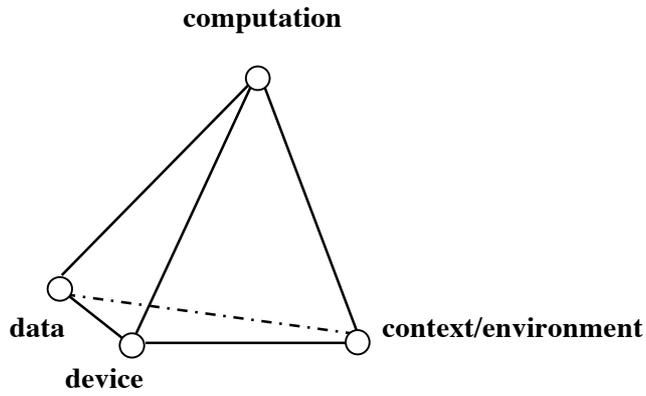

c)   The third face

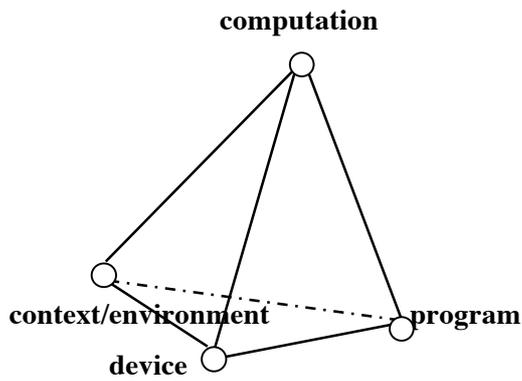

d)   The fourth face

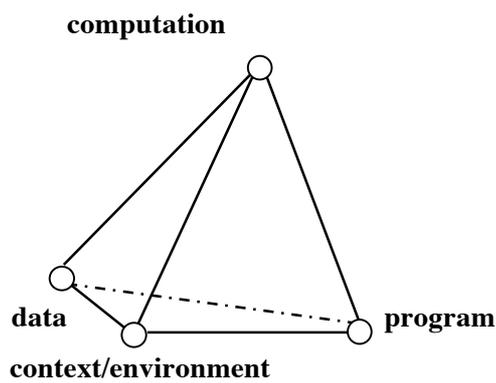

e)   The fifth face

**Figure 11.** The Superspatial Action Computational Pyramid (SSACP)

Exploring the external structure of computation in the context of information processing, we come to the Basic Information Processing Triad Presented in Figure 12. It encompasses three basic information processes (Burgin, 2005).

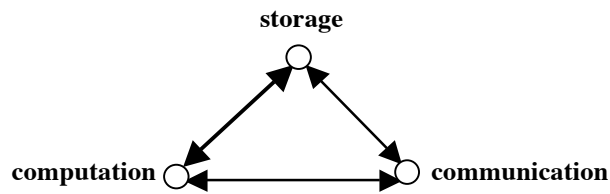

**Figure 11.** The Basic Information Processing Triad (BIPT)

## 5    Natural Computation/ Computing Nature

In the previous sections we presented the development of the concept of computation and its models and developed computational typologies. Adopting the approach of structural realism we provided structural framework of computation based on Burgin's existential triad (physical, structural, mental/cognitive) world of (Burgin, 2012). We have chosen triadic structures as fundamental elements in our study of structures, based on classical existential triad of Plato – (matter, structure, mind). In what follows we will concentrate on the remaining two elements of the existential triad – *physical* and *cognitive*, both of them addressed through the idea of *computing nature/ natural computing*. (Zenil, 2012)(Dodig-Crnkovic & Giovagnoli, 2013)

According to the Handbook of Natural Computing (Rozenberg, Bäck, & Kok, 2012) natural computing is "the field of research that investigates both human-designed computing inspired by nature and computing taking place in nature." In particular, it addresses: intrinsic computation performed by natural materials and computational nature of processes taking place in nature, including living organisms computational as well as models inspired by the natural systems. Natural computing comprises among others areas of cellular automata and neural computation, evolutionary computation, molecular computation and quantum computation, organic computing, biocomputing, nature-inspired algorithms and variety of similar alternative models of computation.

An important characteristic of the research in natural computing is that knowledge is generated bi-directionally, through the interaction between computer science and natural sciences. As the natural sciences are promptly absorbing concepts, tools and methodologies of information processing, computer science is broadening the notion of computation, recognizing information processing found in nature as special type of computation – intrinsic[7], natural computation. (Rozenberg & Kari, 2008) (Stepney et al., 2005)(Stepney et al., 2006) This development in understanding of computation led Denning to argue that computing today is a natural science. (Denning, 2007) Deutsch provided the following definition of computing that encompasses *any physical system* that either computes intrinsically or is used in designed computing machinery:

> "A computing machine is any physical system whose dynamic evolution takes it from one of a set of "input" states to one of a set of "output" states [...] the machine is prepared in a state with a given input label and then, following some motion, the output state is measured. For a classical deterministic system the measured output label is a definite function $f$ of the prepared input label. " (Deutsch, 1985) (p. 97)

---

[7] For more details on the distinction between *intrinsic* and *designed* computation, see (Crutchfield et al., 2010).

In his "Simulating Physics with Computers" Feynman argued that the specific forms of computation are best carried out by the physical substrate we are trying to describe (Feynman, 1982). Even though DNA computation can be *emulated* in silico, the efficiency of computing with DNA substrate instead of its digital representation is higher by many orders of magnitude (Nadin, 2010). Similar view of physical (material) computation is presented in (Stepney, 2008) where it is argued for clear benefits of intrinsic (material) computation, as in the Brook's famous observation that "the world is its own best model" (Brooks, 1990). Cooper addresses the question of the relationship between abstract mathematical models and physical computation that underlies every real-world computation, suggesting that we are under the spell of the "mathematician bias" and arguing for the return to embodied computation (Cooper, Löwe, & Sorbi, 2008; Cooper, 2012a, 2012b) showing how nature can help us to compute. [8]

### 5.1 Development of new models of computation. Intrinsic vs. designed computation

The development of the field of computing, both computing machinery and its models, continues. We are used to quick increase of computational power, memory and usability of our computers, but the limit of miniaturization is approaching as we are getting close to quantum dimensions of hardware components. We are also facing the explosive increase in the amounts of data, "big data" and the emergence of the "internet of things" where computers are becoming an integral part of practically all devices and indeed of our physical environment. All of this poses huge demands on effective computation techniques. Ever since the time of Turing, one of the ideals was *intelligent computing*, which would besides mechanical symbol manipulation include even intelligent problem solving. That would help us manage complexity and vast amounts of data that have to be processed, often in real time. In that direction currently, there is a development of *cognitive computing* (Wang, 2007) aimed towards human level abilities of machines that process/organize/understand information, such as announced by IBM[9].

---

[8] Another take on the same issue is the analysis of the relationship between *axiomatic models* (representing "pure reason") and *construction* (representing empirical contributions to the model development) discussed by Dodig-Crnkovic and Burgin in the chapter *From the Closed Classical Algorithmic Universe to an Open World of Algorithmic Constellations* from (Dodig-Crnkovic & Giovagnoli, 2013).

At the same time the development of computational modelling of human brain has for a goal to reveal the exact mechanisms of human brain function (such as Human Brain Project https://www.humanbrainproject.eu) that will help us understand not only how humans actually perform information processing when they follow an algorithm, *but also how humans create algorithms or models*. Those new developments can be seen as a part of the research within the field of natural computing, where natural system performing computation is human brain.

The new concept of computation with inspiration in natural information processing allows among others learning about nondeterministic complex computational systems, such as living organisms, with self-* properties (self-organization, self-configuration, self-optimization, self-healing, self-protection, self-explanation, and self-awareness). Natural computation has a potential to provide a basis for a unified understanding of phenomena of embodied cognition, intelligence and knowledge generation. (Dodig-Crnkovic & Mueller, 2009)(Wang, 2009)

Recently, a focus issue of the journal Chaos was dedicated to the intrinsic and designed computation under the title "Information Processing in Dynamical Systems—Beyond the Digital Hegemony" addressing challenges of intrinsic computing in dynamical systems as complementary to designed computing in digital systems. (Crutchfield, Ditto, & Sinha, 2010). This relates to the view of a brain as a dynamical system processing information (computing) at different levels of organization – from molecular/electro-chemical, cellular processes (DNA, protein networks, …), neural circuits, cortical columns (morphologically distinct regions of the brain processing and exchanging information), and finally the level of whole-brain information integration that is considered to provide the function of consciousness (Tononi, 2012). Cellular and whole-brain levels of computation correspond to the cognition level of cells and the brain. In a biological sense, cognition in general is the property of an autopoietic system, with self-production, self-organisation and closure and in structural coupling with the environment. (Maturana, 2002)

In the development of new models of computation one path of development goes via unconventional algorithms (Dodig-Crnkovic & Burgin, 2012). New powerful tools that can be used in modelling of *open* systems (such as living systems, including human brains) are brought forth by local mathematics, local logics, logical varieties and the axiomatic theory of algorithms, automata and computation.

At this point we can see several avenues of future development of computing both as physical devices and as models. Natural computation promises new and broader ways of understanding of processes of computation that can be found intrinsically in every physical system. In connection to that, the *architecture* of computational processes becomes increasingly important, especially in the case of cognitive computing.

---

[9] http://www-03.ibm.com/press/us/en/pressrelease/35251.wss

## 5.2 Levels of physical organization in natural computing

In the section on Typology of computation, under the heading Hierarchy of computation levels, we mentioned levels and scales of physical organization at which computation is performed in designed computation. In the case of *The cognitive computing initiative* of IBM[10], for example, the following levels of *designed, nature-inspired* computation are distinguished: synapses, neurons, microcircuits (modelling brains cortical columns function), long range interconnections (modelling long range axons function) and the whole brain level integration of processes, representing *non*-von-Neumann architecture:

> "overarching cognitive computing architecture is an on-chip network of light-weight cores, creating a single integrated system of hardware and software. This architecture represents a critical shift away from traditional von Neumann computing to a potentially more power-efficient architecture that has no set programming, integrates memory with processor, and mimics the brain's event-driven, distributed and parallel processing." http://www-03.ibm.com/press/us/en/pressrelease/35251.wss

When it comes to intrinsic computation, levels of organization at which computation is performed can be connected to different physical forms of natural computers. Quantum computations are performed on the level of quantum-mechanical physical objects. There are many researchers today, who argue that the entire universe (or nature) is a computer or a network of computational processes, among the most prominent ones Lloyd who focus on the universe as quantum computer (Lloyd, 2006). Even Wiesner suggests that nature intrinsically computes on the quantum level:

> "A quantum system can be considered as a computer, storing and processing information during every physical process, be it alignment in spin systems, phase transitions, or chemical bond formation. The amount of matter and energy in the system puts a limit on the amount of computation it can perform and at what speed." (Wiesner, 2010).

In parallel with the developments of natural computing exploring the nature of computation and its limits, experiments are also done with physical properties of materials used for intrinsic computation such as molecular computers that can be in a form of liquid. New materials such as carbon nanotubes are being tested as semiconductors that allow control of electrical signals, instead of silicon that is currently in use. In general, as Cooper and Stepney point out, novel physical sides of computing are becoming increasingly important. (Cooper, 2006)(Stepney, 2008)

---

[10] http://www.hpcwire.com/hpcwire/2011-08-18/ibm_reveals_cognitive_computing_chips.html

## 6      Conclusions and Open Questions

Present account of models of computation first introduces the historical view of scientific concept, suggesting that our idea of computation develops together with the constantly increasing scientific knowledge and tools of analysis (nowadays improved computing machinery helps develop even more advanced computers)(Dodig-Crnkovic, 2013). We then present the structural framework of computation with triadic relationships between (computation, algorithm and device/agent); (data, context/environment and function/goal); (structure, physical and mental/cognitive); (program, device and data), etc. Those are combined to form action computation pyramid with ((data, device/agent, program) and computation).

We highlight several topics of importance for the development of new understanding of computing and its role: natural computation and the relationship between the model and physical implementation, *interactivity* as fundamental for computational modelling of concurrent information processing systems such as living organisms and their networks, and the new developments in modelling needed to support this generalized framework.

The conclusion is that we need much better understanding of computation as information processing than we have now. As there is no information without (physical) representation (Landauer, 1996), the dynamics of information is implemented on different levels of granularity by different physical processes, including the level of computation performed by computing machines (designed computation) and living organisms (intrinsic computation).

There are a still many open problems related to the nature of information and computation, as well as to their relationships. How is information dynamics represented in computational systems, in machines, as well as in living organisms? Are computers processing only data or information and knowledge as well? What do we know of computational processes in machines and living organisms and how these processes are related? What can we learn from natural computational processes that can be useful for information systems and knowledge management?

In sum: the aim of this article is the exposition and delimitation of the possibilities: what do we have today, what (structures and behaviours) can be reached by computational models, and what future developments can be anticipated.

## 7      Acknowledgment

We would like to thank the anonymous referees for constructive comments on the previous version of this article.